# Conformally Invariant Scalar-Vector-Tensor Field Theories Consistent with Conservation of Charge in a Four-Dimensional Space

by

Gregory W. Horndeski
2814 Calle Dulcinea
Santa Fe,NM 87505-6425

e-mail:
horndeskimath@gmail.com

January 24, 2018

# Abstract

In a four-dimensional space I shall construct all of the conformally invariant, scalar-vector-tensor field theories that are consistent with conservation of charge, and flat space compatible. By the last assumption I mean that the Lagrangian of the theory in question, is well defined and differentiable when evaluated for either a flat metric tensor, (and) or constant scalar field, (and) or vanishing vector potential. The Lagrangian of any such field theory can be chosen to be a linear combination of six conformally invariant scalar-vector-tensor Lagrangians, with the coefficients being scalar functions of the scalar field. Five of these generating Lagrangians are at most of second-order, while the sixth one is of third-order. However, the third-order Lagrangian differs from a non-conformally invariant second-order Lagrangian by a divergence. Consequently, all of the conformally invariant scalar-vector-tensor field theories that are consistent with conservation of charge, and flat space compatible, can be obtained from a second-order Lagrangian. The vector equation of any such theory is at most of second-order, and is an extension of Maxwell's equations, incorporating two other first-order terms that vanish when the scalar field is constant. Hence in regions where the scalar field is constant, the vector equation reduces to Maxwell's.

**Section 1: Introduction**

If one is going to use a scalar-tensor field theory to describe gravitational effects in the Universe, then one needs to determine how that theory should be modified to incorporate electromagnetic phenomenon. The simplest way to accomplish this task is to derive the field equations from a Lagrangian L which is the sum of two Lagrangians:

$$L = L_{ST} + L_{SVT}, \qquad \text{Eq.1.1}$$

where $L_{ST}$ is a concomitant of the scalar field $\varphi$, and the gravitational field tensor, $g_{ab}$, along with their derivatives; and $L_{SVT}$ is built from $\varphi$, $g_{ab}$ and the vector potential, $\psi_a$, along with its derivatives. $L_{ST}$ can, *e.g.,* can be chosen from the class of Horndeski Lagrangians presented in [1], which lead to second-order field equations. The possibilities for $L_{SVT}$ are endless. For some guidance in the choice of $L_{SVT}$ let us look at the Einstein-Maxwell equations.

The vacuum field equations of the Einstein-Maxwell field theory are

$$G^{ij} - 2(F^{ia}F^{j}{}_{a} - \tfrac{1}{4}g^{ij}F^{ab}F_{ab}) = 0$$

and

$$F^{ij}{}_{|j} = 0,$$

where the notation employed in this paper is the same as that used in [2] and [3], in terms of which $F_{ab} := \psi_{a,b} - \psi_{b,a}$. A Lagrangian that yields the above field equations is

$$L_{EM} := g^{\frac{1}{2}}R - g^{\frac{1}{2}} F^{ab} F_{ab}. \qquad \text{Eq.1.2}$$

The important thing to note here is that the part of $L_{EM}$ which represents electromagnetic phenomenon; *viz.,* the Maxwell Lagrangian

$$L_M := - g^{\frac{1}{2}} F^{ab} F_{ab}$$

is conformally invariant, and leads to field equations consistent with conservation of charge. In [3] I investigate conformally invariant vector-tensor field theories that are consistent with conservation of charge to determine to what extent $L_M$ can be generalized. If we also require the Lagrangians to be flat space compatible; *i.e.,* such that they are well defined and differentiable when evaluated for either flat metric tensors (and) or vanishing vector fields, then the resulting vector equation must be Maxwell's. Hence we see that the requirements of conformal invariance, conservation of charge and flat space compatibility in a vector-tensor field theory, places quite a severe restriction on the form of the vector field equation. I would now like to investigate what the implications of these restriction are, when they are applied to Lagrangians of the form $L_{SVT}$. To that end I need to introduce some nomenclature to fix ideas.

We shall say that a physical field theory is a SVT (:=scalar-vector-tensor) field theory if the field variables of that theory are the components of a scalar field, $\varphi$, a covariant vector field, $\psi_a$, and a metric tensor field, $g_{ab}$. We shall require the field equations of that theory to be derivable from a variational principle with a Lagrangian of the form

$$L = L(g_{ab}; g_{ab,c}; \ldots; \varphi; \varphi_{,a}; \ldots; \psi_a; \psi_{a,b}; \ldots)  \qquad \text{Eq.1.3}$$

which is scalar density of finite differential order in the field variables. The Euler-Lagrange tensor densities associated with L are defined by

$$E^{ab}(L) := -\frac{\partial L}{\partial g_{ab}} + \frac{d}{dx^c}\frac{\partial L}{\partial g_{ab,c}} + \ldots \qquad \text{Eq.1.4}$$

$$E(L) := -\frac{\partial L}{\partial \varphi} + \frac{d}{dx^a}\frac{\partial}{\partial \varphi_{,a}} + \ldots \qquad \text{Eq.1.5}$$

and

$$E^a(L) := -\frac{\partial L}{\partial \psi_a} + \frac{d}{dx^b}\frac{\partial}{\partial \psi_{a,b}} + \ldots \qquad \text{Eq.1.6}$$

The field theory will be said to be of $k^{th}$ order if one of the sets of field tensor densities has derivatives of at least $k^{th}$ order, in one of the field variables.

Let L be the Lagrangian of a SVT field theory. Under the conformal transformation $g_{ab} \to g'_{ab} := e^{2\sigma}g_{ab}$, where $\sigma$ is a differentiable real valued scalar field, L generates a Lagrangian, L', defined by

$$L'(g'_{ab}; g'_{ab,c}; \ldots; \varphi; \varphi_{,a}; \ldots; \psi_a; \psi_{a,b}; \ldots) :=$$

$$L(g'_{ab}; g'_{ab,c}; \ldots; \varphi; \varphi_{,a}; \ldots; \psi_a; \psi_{a,b}; \ldots).$$

L is said to be conformally invariant if L' = L, when $g'_{ab}$ is replaced throughout L' by $e^{2\sigma}g_{ab}$.

If a SVT field theory is such that $E^a{}_b(L)$, $E^a(L)$ and $E(L)$ are conformally invariant, then that theory will be said to be conformally invariant. If L is conformally invariant, or conformally invariant up to a divergence, then it is well known that its associated SVT field theory will be conformally invariant.

We shall say that a SVT field theory is consistent with conservation of charge, if the Euler-Lagrange tensor density $E^a(L)$ is identically divergence-free. This guarantees that charge is conserved, because in the presence of charge the vector field equation is $E^a(L) = -16\pi J^a$, where $J^a$ is the charge-current vector density. The Einstein-Maxwell field equations generated from the Lagrangian $L_{EM}$ given in Eq.1.2 are consistent with conservation of charge. But they are not conformally invariant, since $E^a_b(L_{EM})$ is not conformally invariant.

Another thing we note about the Einstein-Maxwell field equations is that the Lagrangian of that theory is well defined and differentiable (as a tensorial concomitant) when evaluated for either a flat metric tensor (and) or a vanishing vector field. With that in mind I shall say that a SVT field theory is flat space compatible, if the Lagrangian of that theory is well defined and differentiable when evaluated for either a flat metric tensor, (and) or constant scalar field, (and) or vanishing vector potential. When this is the case the field tensor densities of that theory will also be well defined and differentiable when evaluated for either a flat metric tensor (and) or constant scalar field, (and) or vanishing vector potential. It seems eminently reasonable to demand that a SVT field theory be flat space compatible, since that will guarantee that the Lagrangian, and the field equations of that theory, do not blow up when nothing is in the space.

Since every conformally invariant, flat space compatible ST (:= scalar-tensor)

field theory is trivially consistent with conservation of charge, we can use the work presented in [2] to obtain four classes of Lagrangians which generate SVT field theories which are conformally invariant, consistent with conservation of charge, and flat space compatible. The Lagrangians of those theories are:

$$L_{2C} := g^{½} k(\varphi) \rho^2 , \qquad \text{Eq.1.7}$$

$$L_{3C} := p(\varphi) \varepsilon^{abcd} C^{pq}{}_{ab} C_{pqcd} , \qquad \text{Eq.1.8}$$

$$L_{4C} := g^{½} b(\varphi) C^{abcd} C_{abcd} \qquad \text{Eq.1.9}$$

and

$$L_{UC} := g^{½} u(\varphi)[-12 R^{ab} \varphi_a \varphi_b + 2R\rho - 3(\Box\varphi)^2 - 6\varphi^{ab}\varphi_{ab} - 12\varphi^a \varphi^b{}_{ba}] \qquad \text{Eq.1.10}$$

where $\rho := g^{ab}\varphi_a\varphi_b$, $C^{hijk}$ denotes the components of the Weyl tensor which are defined by

$$C^{hijk} := R^{hijk} + ½(g^{hk}R^{ij} + g^{ij}R^{hk} - g^{hj}R^{ik} - g^{ik}R^{hj}) + \tfrac{1}{6}R(g^{hj}g^{ik} - g^{hk}g^{ij}) , \qquad \text{Eq.1.11}$$

$\varphi_{ab\ldots} := \varphi_{|ab\ldots}$, with k, p, b and u being differentiable functions of $\varphi$, and a vertical bar denoting covariant differentiation. The Euler-Lagrange tensor densities associated with the Lagrangians presented in Eqs.1.7-1.11 can be found in [2].

We shall say that a SVT field theory is a true SVT field theory if all three fields appear somewhere (not necessarily together) in the field equations. Thus the SVT field theories generated by $L_{2C}$, $L_{3C}$, $L_{4C}$ and $L_{UC}$ are not true SVT field theories. However, if we were to add $L_M$ to any combination of those four Lagrangians, we would obtain a true SVT field theory which is conformally invariant, consistent with conservation of charge, and flat space compatible.

Now if we were to multiply the Maxwell Lagrangian by $\beta(\varphi)$, which is an arbitrary scalar function of φ, we would obtain the Lagrangian

$$L_{SM} := -g^{1/2}\beta\, F^{ab}F_{ab} \quad . \qquad \text{Eq.1.12}$$

We could adjoin this Lagrangian to any combination of our four other pure scalar-tensor Lagrangians to obtain true SVT field theories which were conformally invariant, and consistent with conservation of charge. The Euler-Lagrange tensor densities of $L_{SM}$ are given by

$$E^{ab}(L_{SM}) = -2g^{1/2}\beta(F^{ac}F^{b}{}_{c} - \tfrac{1}{4}g^{ab}\,F^{cd}F_{cd}) \qquad \text{Eq.1.13}$$

$$E(L_{SM}) = g^{1/2}\beta' F^{ab}F_{ab}\, , \qquad \text{Eq.1.14}$$

and
$$E^{a}(L_{SM}) = -4g^{1/2}(\beta F^{ab}{}_{|b} + \beta' F^{ab}\varphi_{b})\, , \qquad \text{Eq.1.15}$$

where here " ' " denotes a derivative with respect to φ, and not a conformal transformation.

Associated with the Lagrangian $L_{SM}$ is the Lagrangian $L_{SM^*}$, defined by

$$L_{SM^*} := \gamma\, \varepsilon^{hijk} F_{hi}\, F_{jk} \qquad \text{Eq.1.16}$$

where γ is a differentiable function of φ. Note that when γ is a constant $L_{SM^*}$ is a divergence. The field tensor densities associated with $L_{SM^*}$ are given by

$$E^{ab}(L_{SM^*}) = 0 \qquad \text{Eq.1.17}$$

$$E(L_{SM^*}) = -\gamma'\varepsilon^{hijk}F_{hi}F_{jk}\, , \qquad \text{Eq.1.18}$$

and
$$E^{a}(L_{SM^*}) = 4\gamma'\varepsilon^{aijk}\varphi_{i}F_{jk}\, . \qquad \text{Eq.1.19}$$

It is apparent that $L_{SM^*}$ generates a conformally invariant, flat space compatible SVT field theory, which is consistent with conservation of charge. The purpose of this paper is to demonstrate that in an orientable four-dimensional space, $L_{SM}$ and $L_{SM^*}$ are essentially the only Lagrangians which can be adjoined to $L_{2C}$, $L_{3C}$, $L_{4C}$ and $L_{UC}$ to obtain a SVT field theory with the properties that we are looking for. More exactly, I shall establish the following

**Theorem:** In an orientable four-dimensional pseudo-Riemannian space, any conformally invariant, SVT field theory which is consistent with conservation of charge, and flat space compatible, can have its field equations derived from the Lagrangian

$$L_{2C} + L_{3C} + L_{4C} + L_{UC} + L_{SM} + L_{SM^*} \qquad \text{Eq.1.20}$$

for a suitable choice of the functions k, p, b, u, β and γ appearing in $L_{2C}$, $L_{3C}$, $L_{4C}$, $L_{UC}$, $L_{SM}$ and $L_{SM^*}$ respectively. These six Lagrangians are defined by Eqs.1.7-1.10, 1.12 and 1.16.∎

The first thing you will note about the theorem is that I demand that the spaces of interest must be orientable. This was done to guarantee that the Levi-Civita symbol, $\varepsilon^{abcd}$, is a globally well defined tensor density. However, since most of our work will be done on a coordinate domain, which is an orientable manifold, this assumption is not a severe restriction upon the class of SVT field theories we are investigating. Nevertheless, when we consider coordinate transformations, it will be assumed that the Jacobian is positive.

Another aspect of this theorem that you may have noticed, is that no assumption was made concerning the differential order of the SVT field theories under consideration. That is because I shall prove that all theories which satisfy the assumptions of the theorem must have differential order less that or equal to four. You should also note that all theories satisfying the assumptions of the theorem can be obtained from Lagrangian which is at most of second-order. This is so because the third-order Lagrangian $L_{UC}$ is equivalent to a second-order Lagrangian $L_{2UC}$ (*see,* Eqs.1.25 and 1.26 in [2]) which is conformally invariant up to a divergence.

As an immediate consequence of the Theorem we have the following

**Corollary:** If a SVT field theory satisfies the assumptions of the Theorem, then the Euler-Lagrange tensor density obtained by varying the vector field is given by

$$E^a(L_{SM} + L_{SM*}) = -4g^{½}[\beta F^{ab}{}_{|b} + \beta' F^{ab}\varphi_b - \gamma'\varepsilon^{abcd}\varphi_b F_{cd}] ,  \quad \text{Eq.1.21}$$

for a suitable choice of the scalar functions $\beta=\beta(\varphi)$ and $\gamma=\gamma(\varphi)$. ∎

Eq.1.21 is very interesting. We know that here on earth, Maxwell's equations of electromagnetism do an excellent job of describing electromagnetic effects down to the scale of atoms. Thus if Eq.1.21 is to be taken as a possible generalization of Maxwell's equations, the scalar field must be fairly constant near the earth. In any case, the Corollary tells us that under the assumptions of the theorem there are very few alternatives to Maxwell's equations of electromagnetism.

I shall now quickly sketch how the theorem will be established. I begin by showing that if L satisfies the assumptions of the theorem, then $A^{ij} := E^{ij}(L)$, $B:=E(L)$ and $C^i := E^i(L)$, must be devoid of explicit dependence on the vector field. I then go on to compute what the maximum differential order of $A^{ij}$, B and $C^i$ can be, with $C^i$ turning out to be at most of third-order in $g_{ab}$ and $\varphi$, and second-order in $\psi_a$. Next I construct the general form of $C^i$ and show that $C^i = E^i(L_{SM} + L_{SM*})$. The Lagrangian $L := L - L_{SM} - L_{SM*}$, will satisfy the assumptions of the theorem and be such that $E^i(L) = 0$. The proof then ends by demonstrating that $L$ is equivalent to the Lagrangian $L_{2C} + L_{3C} + L_{4C} + L_{UC}$, for a suitable choice of the functions k, p, b and u. Now for the details, which should be familiar to those who have read [2] and [3].

**Section 2: Proof of the Theorem**

As in [2] and [3] the proof will consist of a sequence of lemmas. Since the signature of the metric tensor will not be significant in what we are about to do, I shall assume that it is arbitrary, but fixed.

The first lemma will provide us with the means to recognize conformally invariant SVT field theories.

**Lemma 1:** Let L be the Lagrangian of SVT field theory. That field theory will be conformally invariant if and only if $E^{ab}(L)$ is trace-free. If $E^{ab}(L)$ is trace-free, then L is conformally invariant up to a divergence.

**Proof:** ⇒The Euler-Lagrange tensor densities of a SVT field theory are related by the identity (*see.,* page 49 of [4])

$$E_a{}^b(L)_{|b} = \tfrac{1}{2}\varphi_a E(L) - \tfrac{1}{2} F_{ab}E^b(L) - \tfrac{1}{2}\psi_a E^b(L)_{|b} \qquad \text{Eq.2.1}$$

If $g'_{ab} := e^{2\sigma}g_{ab}$, we let $E_a{}^b(L)'$, $E(L)'$ and $E^a(L)'$ denote $E_a{}^b(L)$, $E(L)$ and $E^a(L)$ built from $g'_{ab}$, $\varphi$ and $\psi_a$. Since Eq.2.1 is an identity it is valid for every scalar, vector and tensor field. Thus we must have

$$E_a{}^b(L)'_{|'b} = \tfrac{1}{2}\varphi_a E(L)' - \tfrac{1}{2}F_{ab}E^b(L)' - \tfrac{1}{2}\psi_a E^b(L)'_{|'b}, \qquad \text{Eq.2.2}$$

where "$_{|'b}$" denotes covariant differentiation with respect to the Levi-Civita connection of $g'_{ab}$. Due to our assumption of conformal invariance $E(L) = E(L)'$, and $E^a(L) = E^a(L)'$. Since $E^b(L)$ is a contravariant vector density, the right-hand sides of Eq.2.1 and Eq.2.2 are identical. Consequently,

$$E_a{}^b(L)'_{|'b} = E_a{}^b(L)_{|b}. \qquad \text{Eq.2.3}$$

Since, by assumption, $E_a{}^b(L) = E_a{}^b(L)'$, we can use the fact that

$$\Gamma'^r_{st} = \Gamma^r_{st} + \sigma_{,s}\delta^r_t + \sigma_{,t}\delta^r_s - g_{st}\, g^{rm}\sigma_{,m}$$

in Eq.2.3 to deduce that $E_a{}^a(L) = 0$.

⇐Let $g(t)_{ab} := (1-t)g_{ab} + tg'_{ab}$, $0 \leq t \leq 1$, denote the convex combination of $g_{ab}$ and $g'_{ab}$. So $g(t)_{ab} = (1-t + te^{2\sigma})g_{ab}$, is a pseudo-Riemannian metric tensor with the same signature as $g_{ab}$. We now define a one-parameter family of Lagrangians, $L(t)$, by

$$L(t) := L(g(t)_{ab};\, g(t)_{ab,c};\, \ldots;\, \varphi;\, \varphi_{,a};\, \ldots;\, \psi_a;\, \psi_{a,b};\, \ldots).$$

If we let $E^{ab}(L(t))$ denote $E^{ab}(L)$ evaluated for $g(t)_{ab}$, $\varphi$ and $\psi_a$, then it is a straightforward matter to demonstrate that

$$\frac{dL(t)}{dt} = -E^{ab}(L(t))\frac{dg(t)_{ab}}{dt} + \frac{d}{dx^i}V(t)^i, \qquad \text{Eq.2.4}$$

where $V(t)^i$ is a contravariant vector density, built from t, $\sigma$, $g_{ab}$, $\varphi$ and $\psi_a$. Since $E^{ab}(L)$ is trace-free, we know that

$$0 = E^{ab}(L(t))g(t)_{ab} = E^{ab}(L(t)) \, g_{ab}(1-t+te^{2\sigma}) \, ,$$

and thus

$$E^{ab}(L(t)) \frac{dg(t)_{ab}}{dt} = 0 \, .$$

Consequently Eq.2.4 implies that

$$\frac{dL(t)}{dt} = \frac{d}{dx^i} V(t)^i. \qquad\qquad \text{Eq.2.5}$$

If we integrate Eq.2.5 with respect to t from 0 to 1 we find that

$$L(1) - L(0) = \text{a divergence.}$$

But $L(1) = L'$ and $L(0) = L$. Thus if $E^{ab}(L)$ is trace-free, L is conformally invariant up to a divergence. This in turn implies that $E_a{}^b(L)$, $E(L)$ and $E^a(L)$ are conformally invariant.∎

Now that we have found an easy way to recognize conformally invariant SVT field theories, we require an equally facile way to determine when a SVT field theory is consistent with conservation of charge. The next lemma provides us with the means to do just that.

**Lemma 2:** Let $A^{ab}$, B and $C^a$ denote the field tensor densities of a SVT field theory. This theory is consistent with charge conservation if and only if $A^{ab}$, B and $C^a$ are independent of explicit dependence on $\psi_a$.

**Proof:** ⇐Suppose that $A^{ab}$, B and $C^a$ are independent of $\psi_a$; *i.e.,*

$$A^{ab;c} = 0, \; B^{;c} = 0 \; \text{and} \; C^{a;c} = 0,$$

where ";c" denotes a partial derivative with respect to $\psi_c$. Since $A^{ab}$, $B$ and $C^a$ are the field tensor densities of a SVT field theory they must satisfy Eq.2.1, and so,

$$A_a{}^b{}_{|b} = \tfrac{1}{2}\varphi_a B - \tfrac{1}{2}F_{ab}C^b - \tfrac{1}{2}\psi_a C^b{}_{|b}. \qquad \text{Eq.2.6}$$

Upon differentiating Eq.2.6 with respect to $\psi_a$ we get

$$0 = C^b{}_{|b},$$

which implies that charge is conserved.

⇒The proof of the lemma in this direction is virtually identical to the proof of Lemma 2 in [3], and will be omitted.∎

When dealing with SVT field theories that are of $p^{th}$ order in the metric tensor, $q^{th}$ order in the scalar field, and $r^{th}$ order in the vector field, the derivatives of the field tensor densities $A^{ab}$, $B$ and $C^a$ with respect to $\partial^p g_{ab}$, $\partial^q \varphi$ and $\partial^r \psi_a$ are tensorial concomitants. Here I am letting $\partial^p g_{ab}$, $\partial^q \varphi$ and $\partial^r \psi_a$ denote abbreviations for the local components of the $p^{th}$, $q^{th}$ and $r^{th}$ derivatives of $g_{ab}$, $\varphi$ and $\psi_a$. However, in general, the derivatives of $A^{ab}$, $B$ and $C^a$ with respect to $\partial^s g_{ab}$, $s<p$, $\partial^t \varphi$, $t<q$ and $\partial^u \psi_a$, $u<r$; are not tensorial concomitants. But under the assumptions of Lemma 2, $A^{ab;c} = 0$, $B^{;c} = 0$, and $C^{a;c} = 0$, are tensorial equations. Our next lemma addresses the implications of this observation.

**Lemma 3:** In an n-dimensional space, if $A^{ij}$, $B$ and $C^i$ are the field tensor densities of a $k^{th}$ order SVT field theory, which is consistent with conservation of charge, then for every collection of s indicies a, ..., b (s = 2, ..., k+1)

$$\frac{\partial A^{ij}}{\partial \psi_{(a,\ldots b)}} = 0, \quad \frac{\partial B}{\partial \psi_{(a,\ldots b)}} = 0 \text{ and } \frac{\partial C^i}{\partial \psi_{(a,\ldots b)}} = 0,$$

where parentheses around a string of indices denotes symmetrization over the enclosed indices.

**Proof:** The proof is an obvious generalization of the proof of Lemma 3 in [3], and hence it will be omitted. ∎

One might think that Lemma 3 implies that $A^{ij}$, B and $C^i$ must be built from $F_{ab}$ and its derivatives. This is in fact true, and two proofs of this fact are provided in [5]. *E.g.,* in the case where $C^i$ is of fourth-order, we have the replacement theorem

$$C^i = C^i(g_{ab}; \ldots; g_{ab,cdef}; \varphi; \ldots; \varphi_{,abcd}; 0; \tfrac{1}{2}F_{ab}; \tfrac{2}{3}F_{a(b,c)}; \tfrac{3}{4}F_{a(b,cd)}; \tfrac{4}{5}F_{a(b,cde)}).$$

However, we shall not need this fact in what follows.

The next tool we require to prove the Theorem is a generalization of a powerful identity that Aldersley developed to treat conformally invariant concomitants of the metric tensor (*see,* page 70 of Aldersley [6], or [7]). To assist in the statement of this Lemma, I shall use, as I did above, the symbols $\partial^m g_{ab}$, $\partial^m \varphi$ and $\partial^m \psi_a$ as abbreviations for the components of the $m^{th}$ derivatives of $g_{ab}$, $\varphi$ and $\psi_a$.

**Lemma 4: (Aldersley's Lemma for SVT Field Theories)** In an n-dimensional space let

$$A^{ab} = A^{ab}(g_{hi}; \partial g_{hi}; \ldots; \partial^p g_{hi}; \varphi; \partial \varphi; \ldots; \partial^q \varphi; \psi_h; \partial \psi_h; \ldots; \partial^r \psi_h),$$

$$B = B(g_{hi}; \partial g_{hi}; \ldots; \partial^p g_{hi}; \varphi; \partial\varphi; \ldots; \partial^q\varphi; \psi_h; \partial\psi_h; \ldots; \partial^r\psi_h),$$

and

$$C^a = C^a(g_{hi}; \partial g_{hi}; \ldots; \partial^p g_{hi}; \varphi; \partial\varphi; \ldots; \partial^q\varphi; \psi_h; \partial\psi_h; \ldots; \partial^r\psi_h)$$

denote the field tensor densities of a conformally invariant SVT field theory. Then for every real number $\lambda > 0$

$$\lambda^n A^{ab}(g_{hi}; \partial g_{hi}; \ldots; \partial^p g_{hi}; \varphi; \partial\varphi; \ldots; \partial^q\varphi; \psi_h; \partial\psi_h; \ldots; \partial^r\psi_h) =$$

$$A^{ab}(g_{hi}; \lambda\partial g_{hi}; \ldots; \lambda^p\partial^p g_{hi}; \varphi; \lambda\partial\varphi; \ldots; \lambda^q\partial^q\varphi; \lambda\psi_h; \lambda^2\partial\psi_h; \ldots; \lambda^{r+1}\partial^r\psi_h), \quad \text{Eq .2.7}$$

$$\lambda^n B(g_{hi}; \partial g_{hi}; \ldots; \partial^p g_{hi}; \varphi; \partial\varphi; \ldots; \partial^q\varphi; \psi_h; \partial\psi_h; \ldots; \partial^r\psi_h) =$$

$$B(g_{hi}; \lambda\partial g_{hi}; \ldots; \lambda^p\partial^p g_{hi}; \varphi; \lambda\partial\varphi; \ldots; \lambda^q\partial^q\varphi; \lambda\psi_h; \lambda^2\partial\psi_h; \ldots; \lambda^{r+1}\partial^r\psi_h), \quad \text{Eq.2.8}$$

and

$$\lambda^{n-1} C^a(g_{hi}; \partial g_{hi}; \ldots; \partial^p g_{hi}; \varphi; \partial\varphi; \ldots; \partial^q\varphi; \psi_h; \partial\psi_h; \ldots; \partial^r\psi_h) =$$

$$C^a(g_{hi}; \lambda\partial g_{hi}; \ldots; \lambda^p\partial^p g_{hi}; \varphi; \lambda\partial\varphi; \ldots; \lambda^q\partial^q\varphi; \lambda\psi_h; \lambda^2\partial\psi_h; \ldots; \lambda^{r+1}\partial^r\psi) \quad \text{Eq.2.9}$$

where there is no sum over repeated p's, q's or r's in the arguments of $A^{ab}$, $B$ and $C^a$.

**Proof:** The proof is similar to the proof of Aldersley's Lemma given in [2] and [3]. To make the proof more comprehensible, I shall only prove it for the case where $p = q = r = 4$. From that proof it will be evident how to go about establishing the Lemma in general.

Let P be an arbitrary point in our n-dimensional space, and let x be a chart at P. We define a new chart x' at P by $x'^i = \lambda x'^i$. Since $A^{ab}$ is a tensor density we know from the tensor transformation law it must satisfy that at P

$$|\det(J^u{}_v)| J'^a{}_c J'^b{}_d A^{cd}(g_{hi}; \ldots; g_{hi,jklm}; \varphi; \ldots; \varphi_{,jklm}; \psi_h; \ldots; \psi_{h,jklm}) =$$

$$= A^{ab}(g'_{hi}; \ldots; g'_{hi,jklm}; \varphi'; \ldots; \varphi'_{,jklm}; \psi'_h; \ldots; \psi'_{h,jklm}), \quad \text{Eq.2.10}$$

where the Jacobian matrices are defined by $J^u_v := \frac{\partial x^u}{\partial x'^v}$ and $J'^a_c := \frac{\partial x'^a}{\partial x^c}$. The tensor transformation laws for $g_{hi}$, $\varphi$ and $\psi_h$ tell us that

$$g'_{hi} = \lambda^2 g_{hi}; \; g'_{hi,j} = \lambda^3 g_{hi,j}; \ldots; g'_{hi,jklm} = \lambda^6 g_{hi,jklm};$$

$$\varphi' = \varphi; \; \varphi'_{,j} = \lambda \varphi_{,j}; \ldots; \varphi'_{,jklm} = \lambda^4 \varphi_{,jklm};$$

$$\psi'_h = \lambda \psi_h; \; \psi'_{h,j} = \lambda^2 \psi_{h,j}; \ldots; \psi'_{h,jklm} = \lambda^5 \psi_{h,jklm}.$$

Using the above expressions in Eq.2.10, shows us that for every $\lambda > 0$, and any chart x at P

$$\lambda^{n-2} A^{ab}(g_{hi}; g_{hi,j}; \ldots; g_{hi,jklm}; \varphi; \varphi_{,j}; \ldots; \varphi_{,jklm}; \psi_h; \psi_{h,j}; \ldots; \psi_{h,jklm}) =$$

$$A^{ab}(\lambda^2 g_{hi}; \lambda^3 g_{hi,j}; \ldots; \lambda^6 g_{hi,jklm}; \varphi; \lambda \varphi_{,j}; \ldots; \lambda^4 \varphi_{,jklm}; \lambda \psi_h; \lambda^2 \psi_{h,j}; \ldots; \lambda^5 \psi_{h,jklm}). \quad \text{Eq.2.11}$$

I shall now demonstrate how the assumption of conformal invariance can be employed to rewrite Eq.2.11 in the form of Eq.2.7. To that end let $\gamma_{ab}$, $\zeta$ and $\xi_a$ be the x components of a metric tensor, scalar field and vector field defined on a neighborhood of P. Under the conformal transformation $\gamma_{ab} \to \gamma'_{ab} := \lambda^2 \gamma_{ab}$ we find that

$$A^{ab}(\lambda^2 \gamma_{hi}; \lambda^2 \gamma_{hi,j}; \ldots; \lambda^2 \gamma_{hi,jklm}; \zeta; \ldots; \zeta_{,jklm}; \xi_h; \ldots; \xi_{h,jklm}) =$$

$$= \lambda^{-2} A^{ab}(\gamma_{hi}; \gamma_{hi,j}; \ldots; \gamma_{hi,jklm}; \zeta; \ldots; \zeta_{,jklm}; \xi_h; \ldots; \xi_{h,jklm}). \quad \text{Eq.2.12}$$

We set

$$\gamma_{hi} := g_{hi}(P) + \lambda g_{hi,j}(P)\chi^j + \tfrac{1}{2}\lambda^2 g_{hi,jk}(P)\chi^j\chi^k + \tfrac{1}{3!}\lambda^3 g_{hi,jkl}(P)\chi^j\chi^k\chi^l + \tfrac{1}{4!}\lambda^4 g_{hi,jklm}(P)\chi^j\chi^k\chi^l\chi^m,$$

$$\zeta := \varphi(P) + \lambda \varphi_{,j}(P)\chi^j + \tfrac{1}{2}\lambda^2 \varphi_{,jk}(P)\chi^j\chi^k + \tfrac{1}{3!}\lambda^3 \varphi_{,jkl}(P)\chi^j\chi^k\chi^l + \tfrac{1}{4!}\lambda^4 \varphi_{,jklm}(P)\chi^j\chi^k\chi^l\chi^m,$$

and

$$\xi_h := \lambda \psi_h(P) + \lambda^2 \psi_{h,j}(P)\chi^j + \tfrac{1}{2}\lambda^3 \psi_{h,jk}(P)\chi^j\chi^k + \tfrac{1}{3!}\lambda^4 \psi_{h,jkl}(P)\chi^j\chi^k\chi^l + \tfrac{1}{4!}\lambda^5 \psi_{h,jklm}(P)\chi^j\chi^k\chi^l\chi^m,$$

where $\chi^j := x^j - x^j(P)$. Since $\gamma_{hi}(P) = g_{hi}(P)$, $\gamma_{hi}$ is a well defined metric tensor on a neighborhood of P. Using the above expressions for $\gamma_{hi}$, $\zeta$ and $\xi_h$ in Eq.2.12 we find that at P

$$A^{ab}(\lambda^2 g_{hi}; \lambda^3 g_{hi,j}; \ldots; \lambda^6 g_{hi,jklm}; \varphi; \lambda\varphi_{,j}; \ldots; \lambda^4\varphi_{,jklm}; \lambda\psi_h; \lambda^2\psi_{h,j}; \ldots; \lambda^5\psi_{h,jklm}) =$$
$$= \lambda^{-2} A^{ab}(g_{hi}; \lambda g_{hi,j}; \ldots; \lambda^4 g_{hi,jklm}; \varphi; \lambda\varphi_{,j}; \ldots; \lambda^4\varphi_{,jklm}; \lambda\psi_h; \lambda^2\psi_{h,j}; \ldots; \lambda^5\psi_{h,jklm}).  \text{Eq.2.13}$$

Upon combining Eqs.2.12 and 2.13, we discover that at P

$$\lambda^n A^{ab}(g_{hi}; g_{hi,j}; \ldots; g_{hi,jklm}; \varphi; \varphi_{,j}; \ldots; \varphi_{,jklm}; \psi_h; \psi_{h,j}; \ldots; \psi_{h,jklm}) =$$
$$= A^{ab}(g_{hi}; \lambda g_{hi,j}; \ldots; \lambda^4 g_{hi,jklm}; \varphi; \lambda\varphi_{,j}; \ldots; \lambda^4\varphi_{,jklm}; \lambda\psi_h; \lambda^2\psi_{h,j}; \ldots; \lambda^5\psi_{h,jklm}). \quad \text{Eq.2.14}$$

Since P was an arbitrary point, Eq.2.14 is valid in general, and Eq.2.14 agrees with Eq.2.7 when $p = q = r = 4$.

It should be apparent how the above argument can be generalized to demonstrate the validity of Eq.2.7 when p, q and r are arbitrary. In a similar way we can corroborate the validity of Eqs.2.8 and 2.9. ∎

As an immediate consequence of Aldersley's identity we have

**Lemma 5:** In an n-dimensional space, let

$$A^{ab} = A^{ab}(g_{hi}; \partial g_{hi}; \ldots; \partial^p g_{hi}; \varphi; \partial\varphi; \ldots; \partial^q\varphi; \psi_h; \partial\psi_h; \ldots; \partial^r\psi_h)$$
$$B = B(g_{hi}; \partial g_{hi}; \ldots; \partial^p g_{hi}; \varphi; \partial\varphi; \ldots; \partial^q\varphi; \psi_h; \partial\psi_h; \ldots; \partial^r\psi_h)$$

and
$$C^a = C^a(g_{hi}; \partial g_{hi}; \ldots; \partial^p g_{hi}; \varphi; \partial\varphi; \ldots; \partial^q\varphi; \psi_h; \partial\psi_h; \ldots; \partial^r\psi_h),$$

denote the field tensor densitites of a conformlly invariant, flat space compatible, SVT field theory. Then $p \leq n$, $q \leq n$, $r \leq n-1$ in $A^{ab}$ and B, while $p \leq (n-1)$, $q \leq (n-1)$ and $r \leq (n-2)$ in $C^a$. In particular, in a four-dimensional space p and q are $\leq 4$ and $r \leq 3$ in $A^{ab}$ and B, while p and q $\leq 3$, and $r \leq 2$ in $C^a$.

**Proof:** If we differentiate Eq.2.9 with respect to $\partial^r \psi_h$ we obtain

$$\lambda^{(n-1)} \frac{\partial C^a}{\partial(\partial^r \psi_h)}(g_{hi}; \partial g_{hi}; \ldots; \partial^p g_{hi}; \varphi; \partial\varphi; \ldots; \partial^q\varphi; \psi_h; \partial\psi_h; \ldots; \partial^r\psi_h) =$$

$$= \lambda^{(r+1)} \frac{\partial C^a}{\partial(\partial^r \psi_h)}(g_{hi}; \lambda\partial g_{hi}; \ldots; \lambda^p\partial^p g_{hi}; \varphi; \lambda\partial\varphi; \ldots; \lambda^q\partial^q\varphi; \lambda\psi_h; \lambda^2\partial\psi; \ldots; \lambda^{r+1}\partial^r\psi_h)$$

Upon multiplying this equation by $\lambda^{(1-n)}$ we get

$$\frac{\partial C^a}{\partial(\partial^r \psi_h)}(g_{hi}; \partial g_{hi}; \ldots; \partial^p g_{hi}; \varphi; \partial\varphi; \ldots; \partial^q\varphi; \psi_h; \partial\psi_h; \ldots; \partial^r\psi) =$$

Eq.2.15

$$= \lambda^{(r-n+2)} \frac{\partial C^a}{\partial(\partial^r \psi_h)}(g_{hi}; \lambda\partial g_{hi}; \ldots; \lambda^p\partial^p g_{hi}; \varphi; \lambda\partial\varphi; \ldots; \lambda^q\partial^q\varphi; \lambda\psi_h; \lambda^2\partial\psi_h; \ldots; \lambda^{r+1}\partial^r\psi_h).$$

Now if $r-n+2 \geq 1$, then when we take the limit as $\lambda \to 0^+$ in Eq.2.15 the right-hand side vanishes due to flat space compatibility. Therefore if $r \geq n-1$, $C^a$ must be independent of $\partial^r \psi_h$. Consequently if $C^a$ is of $r^{th}$ order in $\psi_h$, $r \leq n-2$.

In a similar way we can establish the other restrictions on p, q and r in $A^{ab}$, B and $C^a$. ∎

Our next objective is to construct all $C^a$'s that satisfy the assumptions of the Theorem. To assist in that endeavor I need to introduce some more notation. If $T^{\cdots}_{\cdots}$ denotes the components of a SVT concomitant we denote the derivatives of $T^{\cdots}_{\cdots}$ with respect to $g_{ab,c\ldots}$; $\varphi_{,ab\ldots}$ and $\psi_{a,b\ldots}$ by

$$T^{\cdots}_{\cdots}{}^{;ab,c\ldots} \; ; \; T^{\cdots}_{\cdots}{}^{;ab\ldots} \text{ and } T^{\cdots}_{\cdots}{}^{;a,b\ldots} \text{ with } T^{\cdots}_{\cdots}{}' := \frac{\partial T^{\cdots}_{\cdots}}{\partial \varphi}.$$

So, *e.g.*,
$$A^{ab;cd,ef} = \frac{\partial A^{ab}}{\partial g_{cd,ef}}, \quad A^{ab:c} = \frac{\partial A^{ab}}{\partial \varphi_{,c}} \quad \text{and} \quad A^{ab;c,d} = \frac{\partial A^{ab}}{\partial \psi_{c,d}}.$$

With this notation in hand I can now state

**Lemma 6:** If $C^a$ satisfies the assumptions of the Theorem then

$$C^{(a;|bc|,def)} = 0, \; C^{(a;bcd)} = 0, \; C^{(a;|b|,cd)} = 0, \; C^{a;b(c,def)} = 0, \; g_{bc}C^{a;bc,def} = 0, \quad \text{Eq.2.16}$$

$$C^{a;(b,c)} = 0 \quad \text{and} \quad C^{a;(b,cd)} = 0. \quad \text{Eq.2.17}$$

**Proof:** From Lemma 5 we know that $C^a$ is at most of third-order in $g_{ab}$ and $\varphi$, and at most of second-order in $\psi_a$. Thus the charge conservation equation $C^a{}_{,a} = 0$, can be written as follows:

$$0 = C^{a;bc}g_{bc,a} + C^{a;bc,d}g_{bc,da} + C^{a;bc,de}g_{bc,dea} + C^{a;bc,def}g_{bc,defa} + C^{a}\varphi_{,a} + C^{a:b}\varphi_{,ba} + C^{a:bc}\varphi_{,bca} +$$
$$+ C^{a:bcd}\varphi_{,bcda} + C^{a;b}\psi_{b,a} + C^{a;b,c}\psi_{b,ca} + C^{a;b,cd}\psi_{b,cda}.$$

Upon differentiating this equation with respect to $g_{rs,tuvw}$, $\varphi_{,rstu}$ and $\psi_{r,stu}$ we obtain

$$0 = C^{(t;|rs|,uvw)}, \; 0 = C^{(r:stu)} \quad \text{and} \quad 0 = C^{(s;|r|,tu)},$$

which establishes the first three conditions in Eq.2.16.

Since $C^a$ is a contravariant vector density it must satisfy various invariance identities (*see, e.g.,* Lovelock and Rund [8]), which can be established as follows.

Let P be an arbitrary point in our space, and let x and x' be charts at P. Due to the tensor transformation law we must have

$$C^a(g'_{hi}; g'_{hi,j}; g'_{hi,jk}; g'_{hi,jkl}; \varphi'; \varphi'_{,j}; \varphi'_{,jk}; \varphi'_{,jkl}; \psi'_h; \psi'_{h,j}; \psi'_{h,jk}) =$$
$$= |\det(J^u{}_v)| J'^a{}_b C^b(g_{hi}; g_{hi,j}; g_{hi,jk}; g_{hi,jkl}; \varphi; \varphi_{,j}; \varphi_{,jk}; \varphi_{,jkl}; \psi_h; \psi_{h,j}; \psi_{h,jk}), \quad \text{Eq.2.18}$$

where the Jacobian matrices $J^u_v$ and $J'^a_b$ have been previously defined. At the point P

$$g'_{hi,jkl} = g_{mn}J^m_{hjkl}J^n_i + g_{mn}J^m_h J^n_{ijkl} + \text{(terms independent of } J^{rs}_{tuv}\text{)},$$

where the $J_{...}$'s are defined inductively by $J^a_{b...cd} := \dfrac{\partial}{\partial x'^d} J^a_{b...c}$. Using this equation, we discover that if we differentiate Eq.2.18 with respect to $J^r_{stuv}$, and evaluate the result for the identity coordinate transformation, we obtain

$$C^{a;hi,jkl}[g_{ri}\delta^s_{(h}\delta^t_j\delta^u_k\delta^v_{l)} + g_{hr}\delta^s_{(i}\delta^{tj}\delta^u_k\delta^v_{l)}] = 0,$$

which implies that

$$C^{a;r(s,tuv)} = 0.$$

Thus we have established the fourth condition in Eq.2.16.

To obtain the last condition of Eq.2.16 let's consider the conformal transformation $g_{ab} \to g'_{ab} := e^{2\sigma}g_{ab}$. $C^a$ is invariant under this transformation, so we must have

$$C^a((e^{2\sigma}g_{hi}); \ldots ; ((e^{2\sigma}g_{hi})_{,jkl}; \varphi; \ldots ; \varphi_{,jkl}; \psi_h; \ldots ; \psi_{h,jk}) =$$
$$= C^a(g_{hi}; \ldots ; g_{hi,jkl}; \varphi; \ldots ; \varphi_{,jkl}; \psi_h; \ldots ; \psi_{h,jk}).$$

If we differentiate this identity with respect to $\sigma_{,rst}$, and then evaluate the result for the identity conformal transformation we obtain

$$C^{a;hi,rst}g_{hi} = 0.$$

This completes our proof of Eq.2.16.

Eq.2.17 follows from Lemma 3. ∎

At last we are ready to determine the basic functional form of $C^a$. This will be done in our next Lemma.

**Lemma 7:** If $C^a$ satisfies the assumptions of the Theorem, then

$$C^a = \Theta_1^{abcdef}g_{bc,def} + \Theta^{abcdefhi}g_{bc,de}\,g_{fh,i} + \Theta_2^{abcdef}g_{bc,de}\,\varphi_{,f} + \Theta^{abcdefhijk}g_{bc,d}\,g_{ef,h}\,g_{ij,k} +$$

$$+\Theta_3^{abcdef}g_{bc,d}\,\varphi_{,e}\,\varphi_{,f} + \Phi_1^{abcd}\varphi_{,bcd} + \Phi_2^{abcd}\varphi_{,bc}\varphi_{,d} + \Phi^{abcdef}\varphi_{,bc}\,g_{de,f} + \Phi_3^{abcd}\varphi_{,b}\varphi_{,c}\varphi_{,d} +$$

$$+\Psi_1^{abcd}\psi_{b,cd} + \Psi^{abcdef}\psi_{b,c}\,g_{de,f} + \Psi_2^{abcd}\psi_{b,c}\varphi_{,d}\,,\qquad\qquad \text{Eq.2.19}$$

where the $\Theta$'s, $\Phi$'s and $\Psi$'s are concomitants of $g_{ab}$ and $\varphi$. $\Theta_1^{abcdef}$, $\Phi_1^{abcd}$ and $\Psi_1^{abcd}$ must have the following symmetries:

$$\Theta_1^{abcdef} = \Theta_1^{a(bc)def} = \Theta_1^{abc(def)},\ \ \Theta_1^{ab(cdef)} = 0\,,\ \ \Theta_1^{(a|bc|def)} = 0,\ g_{bc}\,\Theta_1^{abcdef} = 0\,,\qquad \text{Eq.2.20}$$

$$\Phi_1^{abcd} = \Phi_1^{a(bcd)}\,,\ \ \Phi_1^{(abcd)} = 0\,,\qquad\qquad \text{Eq.2.21}$$

and

$$\Psi_1^{abcd} = \Psi_1^{ab(cd)},\ \ \Psi_1^{(a|b|cd)} = 0\,,\ \ \Psi_1^{a(bcd)} = 0\,.\qquad\qquad \text{Eq.2.22}$$

**Proof:** Due to Lemma 2, Aldersley's Identity (Lemma 4), and Lemma 5, we know that for every $\lambda > 0$,

$$\lambda^3 C^a(g_{hi};\ldots;g_{hi,jkl};\varphi;\ldots;\varphi_{,jkl};\psi_{h,j};\psi_{h,jk}) =$$

$$= C^a(g_{hi};\lambda g_{hi,j};\lambda^2 g_{hi,jk};\lambda^3 g_{hi,jkl};\varphi;\lambda\varphi_{,j};\lambda^2\varphi_{,jk};\lambda^3\varphi_{,jkl};\lambda^2\psi_{h,j};\lambda^3\psi_{h,jk})\,.\qquad \text{Eq.2.23}$$

Upon differentiating this equation with respect to $g_{rs,tuv}$ we find that

$$C^{a;rs,tuv}(g_{hi};\ldots;g_{hi,jkl};\varphi;\ldots;\varphi_{,jkl};\psi_{h,j};\psi_{h,jk}) =$$

$$= C^{a;rs,tuv}(g_{hi};\lambda g_{hi};\lambda^2 g_{hi,jk};\lambda^3 g_{hi,jkl};\varphi;\lambda\varphi_{,j};\lambda^2\varphi_{,jk};\lambda^3\varphi_{,jkl};\lambda^2\psi_{h,j};\lambda^3\psi_{h,jk})\,.\qquad \text{Eq.2.24}$$

If we differentiate this equation with respect to $g_{hi,jkl}$, and then take the limit as $\lambda\to 0^+$, recalling that $C^a$ is well defined and differentiable for a flat metric tensor, constant vector field and vanishing vector potential, we see that

$$C^{a;rs,tuv;hi,jkl} = 0 \ .$$

Similarly we can use Eq.2.24 to prove that

$$C^{a;rs,tuv;hi,jk} = 0 \ , \ C^{a;rs,tuv;hi,j} = 0 \ , \ C^{a;rs,tuv;h,ij} = 0 \ , \ C^{a;rs,tuv;h,i} = 0 \ ,$$

$$C^{a;rs,tuv;j} = 0 \ , \ C^{a;rs,tuv;jk} = 0 \ \text{ and } \ C^{a;rs,tuv;jkl} = 0 \ .$$

Consequently, $g_{bc,def}$ must appear linearly in $C^a$, with coefficients that are functions of only $g_{ab}$ and $\varphi$.

Analogously we can demonstrate that $\varphi_{,bcd}$ and $\psi_{b,cd}$ must appear linearly in $C^a$ with coefficients that are functions of only $g_{ab}$ and $\varphi$.

Continuing in this fashion we can use Eq.2.23 to show that $C^a$ must be a linear combination of

$$g_{bc,def}; \ g_{bc,de}\, g_{fh,i}; \ g_{bc,de}\varphi_{,f} \ ; \ g_{bc,d}\, g_{ef,h}\, g_{ij,k} \ ; \ g_{bc,d}\, \varphi_{,e}\, \varphi_{,f} \ ; \ \varphi_{,bcd}$$

$$\varphi_{,bc}\, \varphi_{,d}; \ \varphi_{,bc}\, g_{de,f} \ ; \ \varphi_{,b}\, \varphi_{,c}\, \varphi_{,d} \ ; \ \psi_{b,cd} \ ; \ \psi_{b,c}\, g_{de,f} \ \text{and} \ \psi_{b,c}\, \varphi_{,d}$$

with coefficients which are simply functions of $g_{ab}$ and $\varphi$.

The symmetries satisfied by $\Theta$, $\Phi$ and $\Psi$ in Eqs.2.20-2.22 follow from Lemma 6, along with the symmetries inherent in the partial derivatives with respect to $g_{bc,def}$, $\varphi_{,bcd}$ and $\psi_{b,cd}$. ∎

In order to simplify the form of $C^a$ given in Lemma 7 we need

**Lemma 8 (Thomas's Replacement Theorem for SVT Concomitants):** If $\tau$ is a tensorial concomitant which locally has the form

$$\tau^{...}_{...} = \tau^{...}_{...}(g_{hi}; g_{hi,j}; g_{hi,jk}; g_{hi,jkl}; \varphi; \varphi_{,h}; \varphi_{,hi}; \varphi_{,hij}; \psi_h; \psi_{h,i}; \psi_{h,ij})$$

then the value of τ's components are unaffected if their arguments are replaced as shown below:

$$\tau^{...}_{...} = \tau^{...}_{...}(g_{hi}; 0; \tfrac{1}{3}(R_{hjki} + R_{hkji}); \tfrac{1}{6}(R_{hjki|l} + R_{hkji|l} + R_{hkli|j} + R_{hlki|j} + R_{hlji|k} + R_{hjli|k});$$

$$\varphi; \varphi_h; \varphi_{hi}; \varphi_{(hij)}; \psi_h; \psi_{h|i}; \psi_{h|(ij)} + \tfrac{1}{6}\psi_m(R^m{}_{i\,mhj} + R_j{}^m{}_{hi})) \, . \qquad \text{Eq.2.25}$$

**Proof:** In [2] and [3] I essentially explain why Thomas's Replacement Theorem [9] gives rise to the result presented in Eq.2.25. ∎

Due to Thomas's Replacement Theorem we see that Eq.2.19 reduces to

$$C^a = \Theta_1^{abcdef} R_{bdec|f} + \Theta_2^{abcdef} R_{bdec}\varphi_f + \Phi_1^{abcd}\varphi_{(bcd)} + \Phi_2^{abcd}\varphi_{bc}\varphi_d + \Phi_3^{abcd}\varphi_b\varphi_c\varphi_d +$$

$$+ \, \Psi_1^{abcd}(\psi_{b|cd} + \tfrac{1}{3}\psi_m R_c{}^m{}_{bd}) + \Psi_2^{abcd}\psi_{b|c}\varphi_d \, , \qquad \text{Eq.2.26}$$

where I have made use of the symmetries of Θ, Φ and Ψ.

At first sight you might think that Eq.2.26 must be incorrect since Lemma 2 stipulates that $C^a$ must be independent of explicit $\psi_a$ dependence. However, we need to know something about $\Psi_1^{abcd}$ before we start to panic. This is where our next Lemma comes to the rescue.

**Lemma 9:** If $\Theta_1^{abcdef}$, $\Phi_1^{abcd}$ and $\Psi_1^{abcd}$ are tensorial concomitants of $g_{ab}$ and $\varphi$ which satisfy Eqs.2.20-2.22, then

$$\Theta_1^{abcdef} = 0 \, , \ \Phi_1^{abcd} = 0 \, , \qquad \text{Eq.2.27}$$

and

$$\Psi_1^{abcd} = g^{\frac{1}{2}} \beta (g^{ab}g^{cd} - \tfrac{1}{2}g^{ac}g^{bd} - \tfrac{1}{2}g^{ad}g^{bc}) \qquad \text{Eq.2.28}$$

where $\beta = \beta(\varphi)$. The tensorial concomitants $\Phi_2^{abcd}$, $\Phi_3^{abcd}$ and $\Psi_2^{abcd}$ have the

following symmetries:

$$\Phi_2{}^{abcd} = \Phi_2{}^{a(bc)d}, \quad \Phi_3{}^{abcd} = \Phi_3{}^{a(bcd)}, \quad \Psi_2{}^{a(bc)d} = 0,$$

and are given by

$$\Phi_2{}^{abcd} = g^{\frac{1}{2}}(\tau\, g^{bc}g^{ad} + \zeta(g^{ab}g^{cd} + g^{ac}g^{bd})), \qquad \text{Eq.2.29}$$

$$\Phi_3{}^{abcd} = g^{\frac{1}{2}}\mu(g^{ab}g^{cd} + g^{ac}g^{bd} + g^{ad}g^{bc}), \qquad \text{Eq.2.30}$$

and

$$\Psi_2{}^{abcd} = g^{\frac{1}{2}}\nu(g^{ab}g^{cd} - g^{ac}g^{bd}) + \omega\, \varepsilon^{abcd}, \qquad \text{Eq.2.31}$$

where $\tau = \tau(\varphi)$, $\zeta = \zeta(\varphi)$, $\mu = \mu(\varphi)$, $\nu = \nu(\varphi)$ and $\omega = \omega(\varphi)$.

**Proof:** In [2] and [3], I build quantities with the same symmetries as $\Theta_1{}^{abcdef}$ and show that they vanish. The main tools used to build concomitants such as $\Theta$, $\Phi$ and $\Psi$, are presented in Weyl [10]. There he demonstrates that concomitants such as those we are trying to construct are generated by all suitable products of $g^{..}$'s and $\varepsilon^{....}$'s. The details of how this is accomplished, are presented in Appendix C of [2].∎

Our next Lemma provides us with our long sought general form for $C^a$.

**Lemma 10:** If $C^a$ satisfies the assumptions of the Theorem then

$$C^a = g^{\frac{1}{2}}\mu\, F^{ab}{}_{|b} + g^{\frac{1}{2}}\mu'\, F^{ab}\varphi_b + \omega\, \varepsilon^{abcd}F_{bc}\varphi_d, \qquad \text{Eq.2.32}$$

where $\mu = \mu(\varphi)$ and $\omega = \omega(\varphi)$ are differentiable functions. A Lagrangian that yields $C^a$ as its Euler-Lagrange tensor density when the vector field is varied is $L_{SM} + L_{SM*}$, where $L_{SM}$ and $L_{SM*}$ are defined by Eqs.1.12 and 1.16 with $\beta := -\frac{1}{4}\mu$, and $\gamma := \frac{1}{4}\int\omega d\varphi$.

**Proof:** Under the assumptions of the Theorem it was shown that $C^a$ must have the form given in Eq.2.26. Thus due to Lemma 9 we can conclude that

$$C^a = g^{\frac{1}{2}} J^a + g^{\frac{1}{2}}\mu\, F^{ab}{}_{|b} + g^{\frac{1}{2}}\nu\, F^{ab}\varphi_b + \omega\, \varepsilon^{abcd} F_{bc}\varphi_d\,, \qquad \text{Eq.2.33}$$

where

$$J^a := \alpha_1 R^{ab}\varphi_b + \alpha_2 R\varphi^a + \alpha_3 \varphi^{ab}\varphi_b + \alpha_4 \varphi^a \Box\varphi + \alpha_5 \varphi^a \rho\,. \qquad \text{Eq.2.34}$$

In deriving the expression for the "junk vector," $J^a$ we need to evaluate $\Theta_2{}^{abcdef} R_{bdec}\varphi_f$. In order to do this one does not really need to employ Weyl's results to first determine the form of $\Theta_2{}^{abcdef}$. It is enough to know that $\Theta_2{}^{abcdef}$ must be built from either the product of three $g^{..}$'s or one $g^{..}$ and one $\varepsilon^{....}$. That is how I obtained the first two terms in the expression for $J^a$. The remaining terms were arrived at using the expressions presented in Lemma 9.

Now $C^a$ is supposed to be conformally invariant. It is clear that the second, third and fourth terms appearing on the right-hand side of Eq.2.33 are conformally invariant. Consequently $g^{\frac{1}{2}} J^a$ must also be conformally invariant. Imposing this demand upon Eq.2.34 shows that $\alpha_1=\alpha_2=\alpha_3=\alpha_4=0$, with $\alpha_5$ being an arbitrary scalar function of $\varphi$. Thus $g^{\frac{1}{2}} J^a$ reduces to

$$g^{\frac{1}{2}} J^a = \alpha_5\, g^{\frac{1}{2}} \varphi^a \rho = E^a(-\alpha_5 g^{\frac{1}{2}}\rho\varphi^b\psi_b)\,. \qquad \text{Eq.2.35}$$

The last term in Eq.2.33 also comes from a variational principle since it is easily seen that

$$\omega\, \varepsilon^{abcd} F_{bc}\varphi_d = E^a(L_{SM*})\,, \qquad \text{Eq.2.36}$$

where $L_{SM*}$ is defined by Eq.1.16 with $\gamma := \frac{1}{4}\int \omega\, d\varphi$.

So right now $C^a$ is given by

$$C^a = \alpha_5 g^{\frac{1}{2}} \varphi^a \rho + g^{\frac{1}{2}}\mu\, F^{ab}{}_{|b} + g^{\frac{1}{2}}\nu\, F^{ab}\varphi_b + \omega\, \varepsilon^{abcd} F_{bc}\varphi_d\,. \qquad \text{Eq.2.37}$$

$C^a$ is supposed to be divergence free. Upon taking the divergence of Eq.2.37 we obtain

$$0 = [g^{\frac{1}{2}}\alpha_5'\rho^2 + g^{\frac{1}{2}}\alpha_5(\varphi^a\rho)_a] + g^{\frac{1}{2}}(\mu' - \nu)F^{ab}{}_{|b}\varphi_a \, .$$

The only way that this equation can hold identically is for $\alpha_5 = 0$, and $\nu = \mu'$. When this choice is made we see that Eq.2.37 implies that

$$C^a = E^a(L_{SM} + L_{SM*}) \, ,$$

with $\beta$ and $\gamma$ in Eqs.1.12 and 1.16, chosen so that $\beta := -\frac{1}{4}\mu$ and $\gamma := \frac{1}{4}\int\omega d\varphi$. This observation completes the proof of the Lemma.∎

We are now sufficiently prepared to finish the proof of the Theorem. To that end let L be a Lagrangian satisfying the assumptions of the Theorem. We define $L := L - L_{SM} - L_{SM*}$, where, due to Lemma 10, we know that we can choose $\beta$ and $\gamma$ in $L_{SM}$ and $L_{SM*}$ so that $E^a(L) = 0$. The purpose of our next Lemma is to determine a pure scalar-tensor Lagrangian equivalent to $L$ from a variational point of view.

**Lemma 11:** Suppose that in a four-dimensional space the $k^{th}$ order Lagrangian $L$ generates a conformally invariant, flat space compatible, SVT field theory for which $E^a(L) = 0$. Then

$$E^{ab}(L) = E^{ab}(L_{2C} + L_{3C} + L_{4C} + L_{UC})$$
and
$$E(L) = E(L_{2C} + L_{3C} + L_{4C} + L_{UC})$$

for a suitable choice of the scalar functions $k(\varphi)$, $p(\varphi)$, $b(\varphi)$ and $u(\varphi)$, appearing in $L_{2C}$, $L_{3C}$, $L_{4C}$ and $L_{UC}$, which are defined by Eqs.1.7-1.10.

**Proof:** Let us consider a 1-parameter variation of $\psi_a$ defined by $\psi(t)_a := t\psi_a$, $0 \leq t \leq 1$. Correspondingly we define a 1-parameter family of Lagrangians $L(t)$ by

$$L(t) := L(g_{ab}; \partial g_{ab}; \ldots; \partial^k g_{ab}; \varphi; \partial\varphi; \ldots; \partial^k\varphi; \psi(t)_a; \partial\psi(t)_a; \ldots; \partial^k\psi(t)_a) . \quad \text{Eq.2.38}$$

Note that since $L$ is flat space compatible, the Lagrangian $L(0)$ is a well defined scalar-tensor Lagrangian which generates a conformally invariant field theory which is trivially consistent with conservation of charge. If we now use the usual variational arguments we find that since $E^a(L(t)) = 0$,

$$\frac{dL(t)}{dt} = \frac{dV(t)^i}{dx^i} \quad \text{Eq.2.39}$$

where $V(t)^i$ is a 1-parameter family of contravariant vector fields. Upon integrating Eq.2.39 with respect to t from 0 to 1, we get

$$L(1) - L(0) = \text{a divergence.} \quad \text{Eq.2.40}$$

Since Eq.2.38 tells us that $L(1) = L$, we may use Eq.2.40 to deduce that the SVT field theory generated by $L$, can also be generated a scalar-tensor Lagrangian which is flat space compatible, and gives rise to a conformally invariant field theory. In [2] I show that the field theory generated by such a scalar-tensor Lagrangian can also be generated by a Lagrangian of the form $L_{2C}+L_{3C}+L_{4C}+L_{UC}$, for a suitable choice of the scalar functions $k(\varphi)$, $p(\varphi)$, $b(\varphi)$ and $u(\varphi)$ appearing in these Lagrangians. This observation completes the proof of the Lemma.∎

Due to Lemmas 10 and 11, we know that if L is a Lagrangian that satisfies the assumptions of the theorem, then there exists scalar functions $\beta=\beta(\varphi)$ and $\gamma=\gamma(\varphi)$ for which $L-L_{SM}-L_{SM*}$ generates a theory that could also be obtained from $L_{2C}+L_{3C}+L_{4C}+L_{UC}$ for a suitable choice of the scalar functions $k(\varphi)$, $p(\varphi)$, $b(\varphi)$ and $u(\varphi)$ in these scalar-tensor Lagrangians. Thus the SVT field theory generted by L can also be generated by $L_{2C}+L_{3C}+L_{4C}+L_{UC}+L_{SM}+L_{SM*}$. This is precisely what we have been trying to prove. So, at long last, our proof of the Theorem is complete.∎

It should be noted that Lemma 11 marks the first and only time that I used the assumption that L was defined and differentiable for a vanishing vector field in the proof of the Theorem. I believe that the Theorem can be proved if we replace the current assumption of flat space compatibility, by the weaker demand that the field tensor densities determined by L are defined and differentiable for either a flat metric tensor, (and) or constant scalar field, (and) or vanishing vector field. (These weaker conditions are all that Aldersley's Lemma requires.) However, proving the Theorem under these weaker assumptions will be much more difficult. What one would have to do is actually construct $A^{ab} := E^{ab}(L)$, and $B := E(L)$, when $E^a(L) = 0$. To that end one can use Aldersley's Identity to get the basic form of $A^{ab}$ and B, as we did for $C^a$ in Lemma 7. Then the problem would be to prove that $A^{ab}$ and B are independent of the vector field, and hence are just scalar-tensor concomitants. (To assist in that endeavor one can use the following facts: $E^c(A^{ab}) = 0$, $E^c(B) = 0$ and $A^{ab}{}_{|b} = \frac{1}{2}\varphi^a B$, when $E^a(L)=0$.) Once this task is accomplished, one can use the Theorem established in [2] to finish the proof.

**Section 3: Concluding Remarks**

In the introduction I briefly discussed how one might go about generalizing the Einstein-Maxwell equations to incorporate a scalar field. Let's attempt to do that by considering a SVT field theory obtained from a Lagrangian L of the form

$$L = L_T + L_{SVT} , \qquad \text{Eq.3.1}$$

where $L_T$ is a pure tensor Lagrangian and $L_{SVT}$ is a scalar-vector-tensor Lagrangian. If we require $L_T$ to generate metric field equations which are at most of second-order, then due to Lovelock's work [11], we know that $L_T$ can be taken to be $L_T = g^{½} \kappa R + g^{½} \Lambda$, where $\kappa$ and $\Lambda$ are constants. Now there are multifarious choices for $L_{SVT}$. If we demand that $L_{SVT}$ satisfies the assumptions of the Theorem then

$$L_{SVT} = L_{2C} + L_{3C} + L_{4C} + L_{UC} + L_{SM} + L_{SM*} \qquad \text{Eq.3.2}$$

where there are six arbitrary scalar functions of $\varphi$ appearing on the right-hand side of Eq.3.2. We now want to find reasons to pare away some of the Lagrangians appearing in Eq.3.2.

The Lagrangian $L_{2C}$ is fairly innocuous, and its field equations are at most of second-order, and quite reasonable. On the other hand, $L_{3C}$ and $L_{4C}$ are more problematic.

$L_{3C}$ is a scalar-tensor version of the Chern-Simons [12] Lagrangian, while $L_{4C}$ is a scalar-tensor version of the Lagrangian that yields the Bach tensor [13]. $L_{3C}$

generates a third-order scalar-tensor field theory, while $L_{4C}$ generates a fourth-order scalar-tensor field theory. It is pointed out in Takahashi and Kobayashi [14], that both of these scalar-tensor theories are afflicted by Ostrogradsky [15] type instabilities. In addition, Crisostomi, *et al.*, [16], have shown that $L_{3C}$'s Ostrogradsky singularity gives rise to two ghosts.

Of the four pure scalar-tensor Lagrangians in Eq.3.2, $L_{UC}$ is my favorite, although it has numerous problems. The field equations generated by $L_{UC}$, are presented in Eqs.1.23 and 1.24 in [2]. The equations $E^{ab}(L_{UC}) = 0$, are of second-order in the metric tensor, and third-order in the scalar field, while the equation $E(L_{UC}) = 0$, is of third-order in the metric tensor and of fourth-order in the scalar field. When working in the vacuum with a theory involving only $L_{UC}$ the equation $E(L_{UC}) = 0$, is satisfied identically when $E^{ab}(L_{UC}) = 0$, due to the identity presented in Eq.2.1. Thus, as far as I am concerned, the vacuum theory generated by $L_{UC}$ is a third-order scalar-tensor field theory. However, $L_{UC}$ has numerous problems as far as Ostrogradsky instabilities go.

In [14] Takahashi and Kobayashi show that, due to the conformal invariance of $L_{UC}$, the theory it generates is degenerate and so it does not satisfy the assumptions of Ostrogradsky's Theorem, which pertains to non-degenerate higher order field theories. Thus one is tempted to say that the theory $L_{UC}$ generates is free of Ostrogradsky ghosts. However, Takahashi has informed me "if one defines

Ostrogradsky ghosts by the appearance of linear momentum in the Hamiltonian...then $L_{UC}$ is plagued by ghosts. Since it suffers from that deficiency." This also follows from the work of Achour, *et al.*, in [17], as well as that of Takahashi and Kobayashi in [14].

Takahashi and Kobayashi also discuss in [14] another problem from which the field theories generated by $L_{UC}$ suffer. It appears that these theories exhibit ghost/gradient instabilities under perturbations about a cosmological background. I am not sure if that should be regarded as the straw that broke the Camel's back as far as $L_{UC}$ is concerned. However, there is another much more serious problem afflicting it.

In the final section of [2] I point out how $L_{UC}$ is the sum of cubic, quartic and quintic Horndeski Lagrangians. It is shown in Ezquiaga and Zumalaćarregui [18], Baker, *et al.*, [19], Sakstein and Jain [20], and Creminelli and Vernizzi [21], that because of the observation of two colliding neutron stars on August 17, 2017, gravitational waves must propagate at the speed of light up to one part in $10^{15}$. These articles explain that this implies that any scalar-tensor Lagrangian involving a quintic Horndeski Lagrangian must be excluded from consideration, since such Lagrangians allow the speed of gravitational waves, denoted by $c_g$, to be appreciably less than c. Thus we must dispense with $L_{UC}$ on very significant physical grounds.

I should also mention that the work by Lombriser and Lima, found in [22], lays the groundwork for some of the analysis presented in [18]-[21].

So right now, if we only consider Lagrangians that are devoid of Ostrogradsky instabilities, and are consistent with observation, then the Lagrangian $L_{SVT}$ of Eq.3.2 reduces to

$$L_{SVT} = L_{2C} + L_{SM} + L_{SM*} .\qquad\text{Eq.3.3}$$

Thus due to Eq.3.1 our simplest scalar-vector-tensor generalization of the Einstein-Maxwell Lagrangian would be

$$L_{simple} = g^{\frac{1}{2}}[\kappa R + \Lambda + k(\varphi)\rho^2 + \beta(\varphi)F^{ab}F_{ab}] + \gamma(\varphi)\,\varepsilon^{abcd}F_{ab}F_{cd} .\qquad\text{Eq.3.4}$$

It has been pointed out to me by L.Heisenberg, that Lagrangians similar to those presented in Eq.3.4, with the addition of a potential term, $g^{\frac{1}{2}}V(\varphi)$, have been extensively investigated in the study of cosmological magnetic fields. For an excellent review article dealing with that subject, please see R.Durrer and A.Neronov [23].

The next simplest modification of the Einstein-Maxwell Lagrangian would be a Lagrangian of the form

$$L_{nx\text{-simplest}} = L_{ST} + L_{SM} + L_{SM*} ,\qquad\text{Eq.3.5}$$

where $L_{ST}$ can be taken to be any scalar-tensor Lagrangian which predicts $c_g \approx 1$. Due to [18]-[21], we know that the quadratic, cubic and quartic Horndeski Lagrangians (with $G_4$ independent of $\rho$) give rise to suitable choices for $L_{ST}$. Some of the Beyond Horndeski Lagrangians for which $c_g \approx 1$, are discussed in Ezquiaga and Zumalaćarregui [18], Crisostomi and Koyama [24], and Dima and Vernizzi [25].

A different type of generalization of the Einstein-Maxwell field equations is provided by the Einstein-Yang-Mills field equations, which present us with an example of a gauge-tensor field theory. (*See*, Yang and Mills, [26], for a discussion of their theory.) If we let $\psi^\alpha{}_i$ denote the gauge potentials (where small Greek indices run from 1 to n, where n is the dimension of the gauge (Lie) group, G), then the components of its associated curvature tensor are given by

$$F^\alpha{}_{ij} := \psi^\alpha{}_{i,j} - \psi^\alpha{}_{j,i} - C^\alpha{}_{\beta\gamma}\psi^\beta{}_i\psi^\gamma{}_j,$$

where $C^\alpha{}_{\beta\gamma}$ denotes the structure constants of the Lie algebra, LG, of the gauge group G. (I realize that my Greek and Latin indices are just the opposite of those conventionally employed, but they are consistent with my previous usage.) A Lagrangian that yields the Einstein-Yang-Mills field equations is given by

$$L_{EYM} := g^{½}\kappa R + g^{½}\Lambda + L_{YM} \qquad \text{Eq.3.6}$$

where

$$L_{YM} := g^{½}B_{\alpha\beta}F^\alpha{}_{ij}F^{\beta ij} \qquad \text{Eq.3.7}$$

and $B_{\alpha\beta}$ denotes the components of a symmetric, Ad G invariant bilinear form on LG. (By $B_{\alpha\beta}$ being Ad G invariant I mean that for every $h \in G$, $B_{\alpha\beta} = B_{\mu\nu}Ad^\mu{}_\alpha(h)Ad^\nu{}_\beta(h)$.) The Yang-Mills Lagrangian given in Eq.3.7 is conformally invariant, and by itself it generates a gauge-tensor field theory which is consistent with conservation of gauge-charge, in that $E^a{}_\alpha(L_{YM})_{\|a} = 0$, where

$$E^a{}_\alpha(L_{YM})_{\|a} := E^a{}_\alpha(L_{YM})_{,a} - E^a{}_\beta(L_{YM})C^\beta{}_{\alpha\gamma}\psi^\gamma{}_a .$$

We recover the Einstein-Maxwell theory (with cosmological term) from this gauge-tensor theory by choosing the Lie group G to be $\mathbb{R}$, and then $B_{\alpha\beta}$ has only one component which we take to equal −1.

Now I believe that it should be possible to modify the theory presented in Section 2 using the material presented in Horndeski [27], to establish the following

**Conjecture:** In an orientable four-dimensional space let L be a Lagrangian which generates a conformally invariant, flat space compatible, scalar-gauge-tensor field theory which is consistent with conservation of gauge charge. Then the Euler-Lagrange tensor densities associated with L can also be obtained form the Lagrangian

$$L_{2C} + L_{3C} + L_{4C} + L_{UC} + L_{SYM} + L_{SYM*}$$

where $L_{2C}$, $L_{3C}$, $L_{4C}$ and $L_{UC}$ are defined by Eqs.1.7-1.10,

$$L_{SYM} := g^{\frac{1}{2}} B_{\alpha\beta}(\varphi) F^\alpha{}_{ij} F^{\beta ij}$$

and

$$L_{SYM*} := D_{\alpha\beta}(\varphi)\, \varepsilon^{hijk} F^\alpha{}_{hi} F^\beta{}_{jk} ,$$

with $B_{\alpha\beta}(\varphi)$ and $D_{\alpha\beta}(\varphi)$ being symmetric, Ad G invariant bilinear forms on LG which are differentiable functions of $\varphi$.∎

If the Conjecture can be proved to be true, then one wonders just how useful that result would be. After all, so far the non-Abelian gauge theories being employed apply in regions governed by quantum mechanics. Clearly that is outside of the realm of a classical field theory. *E.g.*, in [28] I have constructed all of the second-order gauge-tensor field theories that are consistent with conservation of gauge charge, and give rise to the Yang-Mills equation in a flat space. However, very few people have found any use for that result yet. But you can never tell what value a purely mathematical result may have for future physicists.

This completes my trilogy of papers dealing primarily with conformally invariant scalar-vector-tensor field theories. If I were still back in academia I would use these papers as the basis of a graduate course in concomitant theory. In this way I could present the formal ideas behind concomitant theory in the context of interesting problems in field theory.

**Acknowledgements**

I would like to thank Drs. M.Zumalaćarregui, M.Crisostomi, K.Takahashi and L.Heisenberg for numerous very informative discussions on the topics dealt with in this paper. I also wish to thank Drs. J.Sakstein, L.Lombriser and F.Vernizzi for assistance in revising my manuscript. These colleagues have certainly tried their best to bring me up to speed on the plethora of things of which I am ignorant. Whether they have succeeded remains an ongoing concern for all of us.